# The use of squeezed states and balanced homodyne for detecting gravitational waves

Y. Ben-Aryeh

Physics Department, Technion-Israel Institute of Technology, Haifa 32000, Israel

e-mail: phr65yb@physics.technion.ac.il

**ABSTRACT**

The possibility of using squeezed states and balanced homodyne detection of gravitational waves is discussed. It is shown that the quantum noise due to high laser intensities in Michelson interferometer for gravitational waves detection can be reduced by sending squeezed vacuum states to the 'dark' port of the interferometer. The present analysis describes photon statistics measurements effects related to quadrature balanced homodyne detection showing the advantage of using this scheme for detecting gravitational waves.

*Keywords*: gravitational waves detection; squeezed states; balanced homodyne detection ; Michelson interferometer .

# 1. The use of balanced homodyne detection method for quadrature measurements

The principal scheme of using balanced homodyne detector can be explained shortly as follows: The signal interferes with a very strong coherent laser beam, referred as L.O. , at a well–balanced 50:50 beam-splitter (BS). We neglect in this analysis the quantum fluctuations of the L.O. . After the optical mixing of the L.O. with the signal, each beam is directed to a photodetector. The photocurrents are proportional to the photon numbers $\hat{n}_1$ and $\hat{n}_2$ of the beams striking each detector. The photocurrents $I_1$ and $I_2$ are measured and subtracted from each other. A straightforward calculation [1-4] shows that the difference in currents is proportional to:

$$\hat{n}_2 - \hat{n}_1 = \beta^*_{LO}\hat{a} + \beta_{LO}\hat{a}^\dagger \qquad . \qquad (1)$$

Here $\hat{a}$ denotes the annihilation operator of the signal. We use here the notation $\beta$ for the L.O. which is used in the balanced homodyne detection to distinguish it from $\alpha$ which represents the coherent state inserted into the Michelson interferometer. In the present balanced homodyne detection method the weak output in the 'dark' output port of the Michelson interferometer is mixed with the L.O. .

Denoting the the phase of $\beta$ by $\chi$ then we get

$$\hat{n}_2 - \hat{n}_1 = |\beta_{LO}|\hat{X}_{\chi+\pi/2} \quad ; \quad \hat{X}_\chi = \hat{a}e^{-i\chi} + \hat{a}^\dagger e^{i\chi} \qquad . \qquad (2)$$

where $\hat{X}_\chi$ is the quadrature operator of the signal. Notice that the phase $\chi$ is different from the phase $\phi$ which has been defined as the phase of the coherent state inserted in the Michelson interferometer. Notice also the present definition of $\hat{X}_\chi$ given in Eq. (2), where a possible common factor of 2 (or $\sqrt{2}$) in the denominator is omitted for mathematical convenience.

Although more general treatments of balanced homodyne detection have been made [1-4] it has been verified that balanced homodyne detection measures the expectation value for the quadratre operator $\langle \hat{X}_\chi \rangle$ and its variance $(\Delta X_\chi)^2 = \langle \hat{X}_\chi^2 \rangle - (\hat{X}_\chi)^2$.

## 2. Reduction of quantum noise by the use of squeezed states

Previous studies [5-8] have analyzed various phenomena related to the reduction of quantum noise in Michelson interferometer by the use of squeezed states. This topic is especially interesting for the possibility of detecting gravitational waves on earth [9-12]. Since the gravitational waves signals on earth are expected to be extremely weak it is of utmost interest to find methods which will improve the signal to noise ratio of such signals.

In Ref. [7] the authors concentrated about the case in which a squeezed vacuum state is injected into one input port and a very strong coherent state is injected into the other input port of a Michelson interferometer. According to the analysis given in Ref. [7], under the special conditions, defined by $\gamma = (\pi/2) + \delta$, where the parameter $\gamma$ parametrizes the splitting ratio to the incoming beam splitter (BS) in port 1 and $\delta$ is an additional very small phase shift, we obtain in the second output of the interferometer a strong coherent state $|\alpha\rangle$ which is approximately equal to the input coherent state. For gravitational waves detection $|\alpha|$ might be larger than $10^8$ so that under these special conditions it exits in output port 2 approximately unchanged by the squeezed vacuum state. When a vacuum state enters in the 'dark' input port, a small perturbation represented by a small phase shift $\delta$ (induced e.g. by a gravitational wave) leads to coherent state exitting the 'dark' output port with a coherent parameter absolute value given by $|\alpha\delta|$, where $\alpha$ is the coherent parameter of the input very strong coherent state. The insertion of a squeezed vacuum state in the input 'dark' port will change, however, the photon statistics properties in the output 'dark' port which are very important for gravitational waves detection and are studied in the present article.

The result for the output of the interferometer $|\psi\rangle_{OUT}$ under the special conditions given in [8,13] is given as:

$$|\psi\rangle_{OUT} = \hat{D}_1\left\{\alpha\delta\left[(\cosh(r)-1-\sinh r \exp(i(\theta-2\phi))\right]\right\}\hat{S}_1(\varsigma)\hat{D}_1(-\alpha\delta)\hat{D}_2(\alpha)|00\rangle \ . \ (3)$$

Here $\alpha = |\alpha|exp(i\phi)$ represents the very strong coherent input state, $\varsigma = r\exp(i\theta)$ is the squeezing parameter of the input squeezed vacuum state, $\delta$ is a small phase shift introduced by a small external perturbation (e.g. by a gravitational wave), $\hat{S}_1$ is the squeezing operator affecting the weak output in port 1, $\hat{D}_1$ and $\hat{D}_2$ are coherent displacement operators for output port 1 and 2, respectively, $\phi$ and $\theta$ are, respectively, the phases of the coherent and squeezing parameters. Equation (3) is almost equivalent to our previous equation in Ref.[7] (Eq. (79)) but the argument of the first displacement operator has an opposite sign relative to that presented by us [7]. Although the correction in the sign made in Eq. (3) is right [8,13], I would like to show here that basic physical effects of noise reduction are not eliminated. Such analysis has been presented in the previous article [8] showing that that the squeezed vacuum state leads to subpoisson photon numbers distribution in the 'dark' output port and in this way improves the signal to noise ratio. We will review shortly in the present Section such analysis and then show in the next Section the possibility of quadrature measurements which can be obtained by using balanced homodyne detection [1-4]. We will show that the variance of the quadrature measurement can be reduced very much relative to that of coherent state and in this way can improve the phase sensitivity of the interferometer.

Under the condition $\theta - 2\phi = 0$ Eq. (3) gives for the output in port 1 :

$$|\psi\rangle_{OUT,1} \equiv |\psi\rangle_1 = \hat{D}_1\{\alpha\delta(e^{-r}-1)\}\hat{S}_1(\varsigma)\hat{D}_1(-\alpha\delta)|0\rangle_1 \quad . \tag{4}$$

By moving the squeezing operator in Eq. (4) to the left we get for the output in port 1:

$$|\psi\rangle_1 = \hat{S}_1(\varsigma)\hat{D}_1\left[\alpha\delta(e^{-r}-1)[\cosh(r)+\sinh(r)]\right]\hat{D}_1(-\alpha\delta)|0\rangle_1 = \hat{S}_1(\varsigma)\hat{D}_1\left[-\alpha\delta e^r\right]|0\rangle_1. \tag{5}$$

The expectation values for the number operator $\langle \hat{n} \rangle$ and the standard deviation $(\Delta n)^2$ of a squeezed coherent state can be taken from Ref. [4], Eq. (2.7.13) and Eq. (3.5.18), respectively, which are given as:

$$\langle \hat{n} \rangle = |\alpha|^2\left(\cosh^2 r + \sinh^2 r\right) - 2|\alpha|^2\cos(\theta-2\phi)\sinh(r)\cosh(r) + \sinh^2 r \quad , \tag{6}$$

$$(\Delta n)^2 = |\alpha|^2[\cosh(4r) - \cos(\theta-2\phi)\sinh(4r)] + 2\sinh^2 r\cosh^2 r \quad . \tag{7}$$

When we apply these equations to the present system we should insert the following changes: a) $|\alpha|^2$ should be exchanged into $|\alpha\delta|^2 e^{2r}$. b) The phase $\theta - 2\phi$ remains to be equal to 0 under the present condition.

We get for the photon number expectation value in the output port 1: :

$$\langle \hat{n}_1 \rangle \cong |\alpha\delta|^2, \tag{8}$$

so that *the gravitational wave signal is not changed.* For $(\Delta n_1)^2$ we get

$$(\Delta n_1)^2 \cong |\alpha\delta|^2 e^{-2r} \tag{9}$$

so that under the condition $\theta - 2\phi = 0$ the quantum noise represented by Eq. (9) is reduced by a factor $e^{-2r}$, while the signal represented by Eq. (8) remains unchanged. In Eqs. (8-9) we have neglected the small terms which are not proportional to the coherent light intensity.

It follows from the use of Eqs. (8-9) that the gravitational wave signal can be increased by using higher laser intensity which is proportional to $|\alpha|^2$ and at the same time the quantum noise will be reduced by increasing the squeezing parameter $r$ [see also[8]]. One should notice that for $r = 1.3$, which is experimentally attainable, 92.5% of the quantum noise is eliminated, or in other words the signal to noise ratio is improved approximately by a factor 13.5 relative to that of a coherent light.

We should notice also that the effect described in the present Section is critically dependent on the phase parameter $\theta - 2\phi$. The quantum noise is increasing quite rapidly with an increase of this parameter. For example for $r = 1.3$ a large increase of the quantum noise is obtained for an increase of $\theta - 2\phi$ beyond $6^0$. For smaller values of $r$ the improvement of signal to noise ratio is smaller but the range of critical angles is somewhat wider.

In addition to the effect which has been analyzed in the present Section (see also [8]) for subpoisson photon number distribution, we show in the following Section the effect of quadrature squeezing. For balanced homodyne detection measurements the weak output in the 'dark' output port is mixed with an additional strong coherent state (L.O.). We will concentrate on the condition $\theta - 2\phi = 0$ for which as shown in the present Section strong subpoisson effects are obtained. We will show that under this condition a very strong quadrature squeezing is obtained leading to a strong phase sensitivity of the interferometer. We will show that the variance of the quadrature measurement can be reduced very much relative to that of a coherent state and in this way can improve the phase sensitivity of the interferometer.

## 3. Quadrature measurements in the output of Michelson interferometer

Classically it is well known [9] that amplification of an optical signal involves a corresponding increase in the amount of noise so that the signal to noise ratio is not improved by the amplification process. Such conclusion is however not valid for nonclassical light in which the amplification is phase dependent [14]. In addition to the effects which are related to subpoisson photon number distribution, we show here certain effects related to quadrature measurements

In the derivations of the present analysis we use the following relations for the squeezing operators [4]:

$$\hat{S}_1^{\dagger}(\varsigma)\hat{a}\hat{S}_1(\varsigma) = \hat{a}\cosh(r) - \hat{a}^{\dagger}\sinh(r)\exp(i\theta) \quad ;$$
$$\hat{S}_1^{\dagger}(\varsigma)\hat{a}^{\dagger}\hat{S}_1(\varsigma) = \hat{a}^{\dagger}\cosh(r) - \hat{a}\sinh(r)\exp(-i\theta) \quad . \tag{10}$$

The displacement operators satisfy [4]:

$$\hat{D}_1^{\dagger}[\alpha]\hat{a}\hat{D}_1[\alpha] = \hat{a} + \alpha$$
$$\hat{D}_1^{\dagger}[\alpha]\hat{a}^{\dagger}\hat{D}_1[\alpha] = \hat{a}^{\dagger} + \alpha^* \tag{11}$$

We can change the order of the diaplacement and squeezing operators using the relation [4]:

$$\hat{D}(\alpha)\hat{S}(\varsigma) = \hat{S}(\varsigma)\hat{D}\left(\alpha\cosh(r) + \alpha^* e^{i\theta}\sinh(r)\right) \quad ; \quad \varsigma = re^{i\theta} \quad . \tag{12}$$

Since our interest is in photon statistics we can use the relation:

$$\hat{D}(\alpha_1)\hat{D}(\alpha_2) = \hat{D}(\alpha_1 + \alpha_2) \quad , \tag{13}$$

where in Eq. (13) we neglect a phase term which does not affect photon statistics.

In the present Section we analyze balanced homodyne method for detecting gravitational waves based on the use of Eq. (3).

Using Eq. (12) we get the relation

$$\hat{D}_1\{\alpha\delta[(\cosh(r)-1-\sinh r \exp(i(\theta-2\phi))]\}\hat{S}_1(\varsigma) \\ = \hat{S}_1(\varsigma)\hat{D}_1\{\alpha\delta[1-\cosh(r)-\sinh(r)\exp(i(\theta-2\phi))]\} \quad . \tag{14}$$

Substituting Eq. (14) into Eq. (3) we get for the wavefunction in output port 1:

$$|\psi\rangle_1 = \hat{S}_1(\varsigma)\hat{D}_1\{-\alpha\delta[\cosh(r)+\sinh(r)\exp(i(\theta-2\phi))]\}|0\rangle_1 \quad . \tag{15}$$

For $\theta - 2\phi = 0$ the result of Eq. (15) is reduced to that of Eq. (5).

Let us define:

$$\alpha' = \{-\alpha\delta[\cosh(r)+\sinh(r)\exp(i(\theta-2\phi))]\} \quad . \tag{16}$$

Then Eq. (15) can be written as

$$|\psi\rangle_1 = \hat{S}_1(\varsigma)\hat{D}_1\{\alpha'\}|0\rangle_1 = \hat{S}_1(\varsigma)|\alpha'\rangle \quad , \tag{17}$$

where $\alpha'$ is the coherent parameter of the coherent state $|\alpha'\rangle$.

For the absolute value of $\alpha'$ we get the relation:

$$|\alpha'|^2 = |\alpha\delta|^2\{\cosh^2(r)+\sinh^2(r)+2\sinh(r)\cosh(r)\cos(\theta-2\phi)\} \quad , \tag{18}$$

so that for large values of $r$ and for $|\theta-2\phi| \leq 90^0$ we get $|\alpha'|^2 \gg |\alpha\delta|^2$. The phase of $\alpha'$ can be calculated from Eq. (16).

The expectation value for the quadrature is given by

$$\langle \hat{X}_\chi \rangle = \langle \alpha' | \hat{S}_1(\varsigma)^\dagger \left( \hat{a} e^{-i\chi} + \hat{a}^\dagger e^{i\chi} \right) \hat{S}_1(\varsigma) | \alpha' \rangle =$$

$$\langle \alpha' | \begin{bmatrix} \left( \hat{a} \cosh(r) - \hat{a}^\dagger \sinh(r) \exp(i\theta) \right) e^{-i\chi} + \\ \left( \hat{a}^\dagger \cosh(r) - \hat{a} \sinh(r) \exp(-i\theta) \right) e^{i\chi} \end{bmatrix} | \alpha' \rangle \qquad (19)$$

$$= \left\{ \left[ \alpha' \cosh(r) - \alpha'^* \sinh r \exp(i\theta) \right] e^{-i\chi} + C.C. \right\}$$

$$= 2|\alpha'| \left[ \cosh(r) \cos(\phi' - \chi) - \sinh(r) \cos(\phi' + \chi - \theta) \right]$$

Here $\phi'$ is the phase of $\alpha'$ and C.C. denotes the complex conjugate of the previous expression. The phase $\phi'$ can be calculated from Eq. (16) depending on the parameters $r$, $\phi$ and $\theta - 2\phi$.

The expectation value for the quadrature squared is given by:

$$\langle \hat{X}_\chi^2 \rangle =$$

$$\langle \alpha' | \hat{S}_1(\varsigma)^\dagger \left( \hat{a} e^{-i\chi} + \hat{a}^\dagger e^{i\chi} \right)^2 \hat{S}_1(\varsigma) | \alpha' \rangle = \langle \alpha' | \begin{bmatrix} \left( \hat{a} \cosh(r) - \hat{a}^\dagger \sinh(r) \exp(i\theta) \right) e^{-i\chi} + \\ \left( \hat{a}^\dagger \cosh(r) - \hat{a} \sinh(r) \exp(-i\theta) \right) e^{i\chi} \end{bmatrix}^2 | \alpha' \rangle =$$

$$\langle \alpha' | \begin{bmatrix} \left( \hat{a}^2 \cosh^2(r) + \hat{a}^{\dagger 2} \sinh^2(r) \exp(i2\theta) - \left(2\hat{a}^\dagger \hat{a} + 1\right) \sinh(r) \cosh(r) \exp(i\theta) \right) e^{-i2\chi} + \\ \left( \hat{a}^{\dagger 2} \cosh^2(r) + \hat{a}^2 \sinh^2(r) \exp(-i2\theta) \right) - \left(2\hat{a}^\dagger \hat{a} + 1\right) \sinh(r) \cosh(r) \exp(-i\theta) e^{i2\chi} + \\ \left(2\hat{a}^\dagger \hat{a} + 1\right) \cosh^2 r + \left(2\hat{a}^\dagger \hat{a} + 1\right) \sinh^2 r - \sinh(r) \cosh(r) [2\hat{a}^2 \exp(-i\theta) + 2\hat{a}^{\dagger 2} \exp(i\theta)] \end{bmatrix} | \alpha' \rangle$$

$$\approx |\alpha'|^2 \begin{bmatrix} 2\cosh^2 r \cos(2\phi' - 2\chi) + 2\sinh^2 r \cos(2\phi' + 2\chi - 2\theta) \\ -4\sinh(r)\cosh(r)\{\cos(2\chi - \theta)\} - 4\sinh(r)\cosh(r)\{\cos(2\phi' - \theta)\} + 2\cosh(2r) \end{bmatrix}$$

$$+ \left[ \cosh(2r) - \sinh(2r) \cos(\theta - 2\chi) \right]$$

. (20)

We find that the following two parameters are needed for calculating $\langle \hat{X}_\chi^2 \rangle$ and $\langle \hat{X}_\chi \rangle$:

$$\xi = \chi - \phi' \quad ; \quad \eta = \theta - 2\phi' \quad . \qquad (21)$$

as the following relations are obtained:

$$(\phi' + \chi - \theta) = \xi - \eta \quad ; \quad 2\chi - \theta = 2\xi - \eta \quad ; \quad (2\phi' + 2\chi - 2\theta) = 2\xi - 2\eta \quad . \quad (22)$$

The analysis for quadrature measurements turn to be quite complicated under general conditions for $\theta - 2\phi$. We are interested, however, in simplifying the analysis under the condition $\theta - 2\phi = 0$ for which subpoisson photons distribution is obtained, as analyzed in the previous Section (see also [8]) According to Eq. (16) by assuming $\theta - 2\phi = 0$ we get $\phi' = \phi - \pi$ and under this condition $\eta = \theta - 2\phi' = \theta - 2\phi = 0$. We find then that for this case the expectation value for the quadrature and the quadrature squared depend only the phase $\xi = \chi - \phi'$. According to Eq. (19) we get :

$$\langle \hat{X}_\chi \rangle = 2|\alpha'|e^{-r}\cos(\xi) \qquad (23)$$

For $\theta - 2\phi = 0$ we get according to Eq. (20):

$$\langle \hat{X}_\chi^2 \rangle \approx |\alpha'|^2 4e^{-2r}\cos^2(\xi) + [\cosh(2r) - \sinh(2r)\cos(2\xi)] \qquad (24)$$

We should take into account that according to Eq. (16) for $\theta - 2\phi = 0$ we get

$$\alpha' = -\alpha\delta e^r \quad , \qquad (25)$$

so that the dependence in Eqs. (23) and (24), respectively, on $e^{-r}$ and $e^{-2r}$ is eliminated and we get:

$$\langle \hat{X}_\chi \rangle = 2|\alpha\delta|\cos(\xi) \quad , \qquad (26)$$

$$\langle \hat{X}_\chi^2 \rangle = |\alpha\delta|^2 4\cos^2(\xi) + [\cosh(2r) - \sinh(2r)\cos(2\xi)] \qquad (27)$$

The quadrature variance is given as

$$\langle \hat{X}_\chi^2 \rangle - \langle \hat{X}_\chi \rangle^2 = [\cosh(2r) - \sinh(2r)\cos(2\xi)] \qquad (28)$$

Using the definition of $\hat{X}_\chi$ given in Eq. (2) we find that while for a coherent state one gets [4]:

$$\langle \hat{X}_\chi^2 \rangle - \langle \hat{X}_\chi \rangle^2 = 1 \quad , (for \quad coherent \quad state) \quad , \tag{29}$$

we get according to Eq. (28) a squeezed state represented by:

$$\langle \hat{X}_\chi^2 \rangle - \langle \hat{X}_\chi \rangle^2 = e^{-2r} \quad , \quad for \quad \xi = 0 \quad , \tag{30}$$

$$\langle \hat{X}_\chi^2 \rangle - \langle \hat{X}_\chi \rangle^2 = e^{2r} \quad , \quad for \quad 2\xi = \pi. \tag{31}$$

According to Eqs. (30-31) we get strong squeezing effects for large values of the squeezing parameter $r$.

While in the previous analysis [8] the gravitational waves signal was amplified only by increasing the value of the coherent state $|\alpha|$, in the present balanced homodyne detection scheme the gravitational waves signal is amplified according to Eqs. [1-2] also by the value of $|\beta|$. From the technical point of view there might be certain problems in increasing the interferometer laser intensity beyond a certain limit (e.g. perturbations by large radiation pressures [5,6]). There is therefore a great advantage in the present scheme in which the gravitational waves signal is increased also by the strong value of $|\beta|$.

Balanced homodyne detection method can be used for observing optical signals with high resolution [3]. In the present analysis it follows from Eq. (2) that the homodyne detection signal is amplified by using higher values of $|\beta|$. The L.O. coherent parameter ($|\beta|$ is, however, assumed to be much smaller than the extreme high value of $|\alpha|$ used in the Michelson interferometer. The fluctuations in the value of the quadrature $\hat{X}_\chi$ (representing 'phase fluctuations' [4]) can be decreased by using an optimal balanced homodyne phase $\xi$ leading to very small quadrature fluctuations given by Eq. (30).

# 4. Discussion and summary

One of the most important challenges of modern physics is the possibility to detect gravitational waves on earth. In spite of the large effort invested in this project [12], so far, this goal has not been achieved. Due to the fact that there are abundant noise sources on earth for gravitational waves with frequencies below 1 cycle/sec (e.g. seismic, thermal, mechanical [15], etc.) the extensive research in this field has been concentrated on gravitational waves in the frequency region 1-1000 cycles/sec. In this gravitational waves frequencies' region the dominant noise factor is the shot noise which is expected to be reduced by using very high laser intensities.

In the present work the use of Michelson interferometer for detecting gravitational waves is studied for the case in which a very strong laser light is inserted in one input port of the interferometer and a squeezed vacuum state is inserted in the other input port of the interferometer. A small phase change induced by a gravitational wave is expected to affect the output in the 'dark' port of the interferometer and photon statistics measurements are expected to be affected by the use of the squeezed vacuum states. Although this physical system has been studied in previous publications [7,8,13] certain important results related to improvement of the signal to noise ratio, especially those related to quadrature measurements, have not been analyzed, and are treated in the present article.

It has been shown in Section 2 (see also [8]) that subpoisson photon statistics is implemented by the use of the squeezed vacuum state under certain critical parameters. The gravitational wave optical signal can be increased by using higher laser intensity while due to the use of squeezed states the coherent quantum noise will not increased, only that given by a small amount of squeezed state. It has been shown in Section 3 that under the critical parameters for which subpoisson photon statistics is obtained strong quadrature squeezing is also obtained. Such quadrature squeezing measurement is implemented by balanced homodyne detection obtaining amplification with very small phase fluctuations.

For emphasizing the importance of balanced homodyne detection scheme we can quote from Ref. [3]:"The balanced homodyne detector … is an amplifier…the balanced homodyne detector takes advantage of the high efficiency of photodiodes and at the same time can determine signal with single photon-resolution-a nearly perfect technical solution!" .